\pdfoutput=1

\documentclass[notitlepage, twoside, 11pt]{report}

\usepackage{caption}
\usepackage{subcaption}
\usepackage{mathrsfs}
\usepackage{nicefrac}
\usepackage{amssymb}
\usepackage{comment}
\usepackage[tight,english]{minitoc}
\usepackage{enumitem}

\usepackage{booktabs}
\usepackage{colortbl}
\usepackage{multirow}

\usepackage{arsclassica}
\usepackage{acronym}

\usepackage{graphicx,bm,amsmath}

\usepackage{hyperref}
\usepackage[a4paper, bottom=1.5in, total={6in, 9in}]{geometry}
\usepackage{tcolorbox}

\usepackage{bbold}

\newcommand{\be}{\begin{equation}}
\newcommand{\ee}{\end{equation}}
\newcommand{\bea}{\begin{eqnarray}}
\newcommand{\eea}{\end{eqnarray}}

\definecolor{smoothred}{HTML}{C5232F}
\definecolor{mygreen}{rgb}{0,0.5,0}
\definecolor{myblue}{rgb}{0,0,0.75}
\definecolor{mymagenta}{cmyk}{0,1,0,0.12}

\newcommand{\samnote}[1]{{\color{gray} #1}}

\usepackage{authblk}

\makeatletter
\newcommand*{\toccontents}{\@starttoc{toc}}
\makeatother

\begin{document}
\pagenumbering{gobble}
\author[1,2]{Samuele Cavinato}

\author[1]{Alessandro Scaggion}

\affil[1]{Medical Physics Department, Veneto Institute of Oncology, IOV-IRCCS, 35128 Padova, Italy}
\affil[2]{Dipartimento di Fisica e Astronomia ``G. Galilei'', Universit\`a di Padova, 35131 Padova, Italy}
\date{\today}
\title{{\Huge{TCoMX v$1.0$}}\\\textit{User manual}}
\date{\today \\ \par\noindent\rule{\textwidth}{0.4pt}}

\maketitle

\toccontents
\par\noindent\rule{\textwidth}{0.4pt}
\clearpage
\pagenumbering{roman}
\chapter*{Abbreviations and definitions}
\begin{acronym}[MLC]
\acro{MLC}{Multi-leaf collimator}
 
\acro{$N_{leaves}$}{Number of leaves}

\samnote{\small It is the number of leaves of the binary MLC.\par}

\acro{LOT}{Leaf Open Time}

\samnote{\small It indicates the time duration of leaf aperture. It is usually measured in milliseconds (ms).\par}

\acro{FLOT}{Fractional Leaf Open Time}

\samnote{\small It indicates the fractional duration of an MLC opening for a given leaf. It is a number in~{$[0,1]$}.\par}

\acro{S}{Sinogram}

\samnote{\small It is an  ${(N_{proj} \times N_{leaves})-}$dimensional matrix containing the FLOT for each leaf (columns) at each projection (rows).\par}

\acro{O}{Mask-sinogram}

\samnote{\small It is an  ${(N_{proj} \times N_{leaves})-}$dimensional matrix whose entries are set to $1$ for all FLOTs  $>0$, $0$ otherwise.\par}

\acro{P}{Leaf position array}

\samnote{\small It is an  ${(1 \times N_{leaves})-}$dimensional array whose entries of mark the central position of each leaf.\par}

\end{acronym}

\clearpage

\pagenumbering{arabic}

\chapter{The TCoMX library} 
The TCoMX (Tomotherapy Complexity Metrics EXtractor) library is a in-house developed Matlab{\textregistered} (The Mathworks Inc, Natick, MA, USA) library for the automatic extraction of  a wide set of complexity metrics from the DICOM RT-plan files of helical tomotherapy (HT) plans (\href{(https://github.com/SamueleCavinato/TCoMX)}{https://github.com/SamueleCavinato/TCoMX}). The current version of {TCoMX (v$1.0$)}  allows the extraction of all the different complexity metrics proposed in the literature, some of them with customisable parameters. \\
{TCoMX} is compatible with DICOM RT-PLAN files generated using both RayStation (RaySearch Laboratories, Stockholm, Sweden) and Precision (Accuray, Sunnyvale, CA) TPSs. It was developed entirely on Matlab{\textregistered}  R2020b. The backward compatibility with previous Matlab{\textregistered}  releases was checked up to version R2018a. Compatibility with older versions should be guaranteed but needs to be verified. The library was developed on Linux Ubuntu 20.04.1 LTS and the compatibility with Windows 10 was verified. The compatibility with other versions of the two operating systems as well as with other operating systems needs to be investigated. 
A reference dataset composed by 18 anonymized DICOM RT-PLAN files (9 Precision, 9 RayStation) is also provided in the repository and it is described in Chapter~\ref{ch:reference_dataset}.
\begin{figure}[ht!]
    \centering
    \includegraphics[width=1.0
    \textwidth]{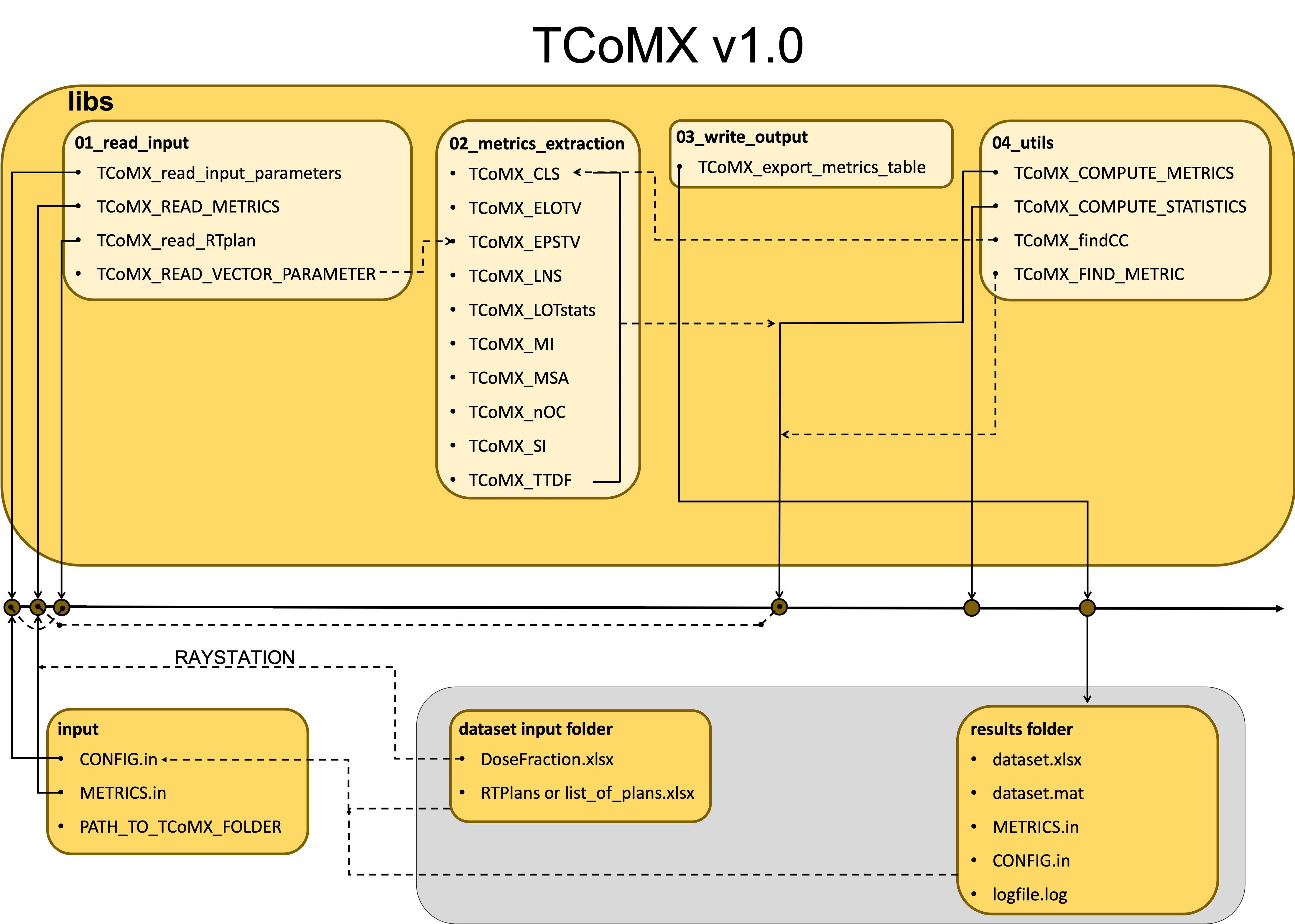}
    \caption{Structure of the TCoMX library. Yellow boxes represent the folders inside the TCoMX-main folder, while the gray box those outside. Dependencies between functions, files or folders are marked using dashed arrows/lines. The flow of the execution is marked by the horizontal arrow in the middle of the figure. Brown circles are placed on the line according to the order the different functions are called during the execution. Solid arrows mark the flow of the inputs/outputs.}
    \label{fig:TCoMX_structure}
\end{figure}
\section{Overview}
In Fig.~\ref{fig:TCoMX_structure}, a schematic of the functioning of the TCoMX library is shown. In the first two stages the library reads the input parameters, such as the input and results directories and the metrics to compute from the dedicated input files $CONFIG.in$ and $METRICS.in$ using the functions TCoMX$\_$read$\_$input$\_$parameters and TCoMX$\_$READ$\_$METRICS. Details concerning the input files are provided in see Sec.~\ref{sec:input_files}. The input directory is needed to the TCoMX$\_$read$\_$RTplan function to read the plans. Metrics are extracted from each plans by calling the TCoMX$\_$COMPUTE$\_$METRICS function, which calls all the functions in the 02$\_$metrics$\_$extraction folder and extracts the metrics according to what specified in the METRICS.in file. The function TCoMX$\_$FIND$\_$METRIC is used to select the metrics to compute according to what reported in the \textit{METRICS.in} file. \\Once the metrics have been extracted from all the plans, results are exported using onto a file called $dataset.xlsx$ which is stored into the results folder specified in the $CONFIG.in$ file using the function TCoMX$\_$export$\_$metrics$\_$table.
\section{Input files}
\label{sec:input_files}
The TCoMX library was developed to be used by any user. For this reason, no knowledge of Matlab coding is required. The interaction of the user with the library is done through the following text files: CONFIG.in and METRICS.in. In a following version, a Graphic User Interface (GUI) will be available. Furthermore, an executable version of the library will be available in the future. 
\subsection{CONFIG.in}
This file is used to set all the necessary input parameters. The overall structure of the file is the following:
{\small{\begin{tcolorbox}[rounded corners, colback=gray!30, colframe=gray!80!, title= \textbf{CONFIG.in}]
\begin{itemize}[leftmargin=*]
    \item[] \textbf{$\#$ Analysis ID name}
    \newline (OPTIONAL) Type here a user's define name to assign to the current session
    \item[] \textbf{$\#$ Dataset input folder}
    \newline Type here the full path where your DICOM RTPLAN files or a file containing the list of plans (see below) are stored.
    \item[] \textbf{$\#$ Results output folder}
    \newline Type here the full path where the results of the extraction process will be saved
    \item[] \textbf{$\#$ Read from file}
    \newline \textit{True} or \textit{False}. If it is set to \textit{true}, the library will read the filenames from a dedicated .xlsx file stored in the the 'Dataset input folder'. The name of the .xlxs file is specified in the 'List file name' field (see below). An example of this file is provided together with the reference dataset.
    \item[] \textbf{$\#$ List file name}
    \newline .xlsx file containing the list of plans. It is used only if 'Read from file' is set to \textit{true}.
\end{itemize}
\end{tcolorbox}}}
    \begin{tcolorbox}[rounded corners, colback=red!30, colframe=red!80!, title= \textbf{WARNING}]
    \begin{itemize}[leftmargin=*]
        \item The fields with the $\#$ must be left as they are.
        \item The value of a given field must be written in the line below the corresponding field name as shown in the CONFIG.in box above. 
        
    \end{itemize}
    \end{tcolorbox}
\subsection{METRICS.in}
This file is used to select the metrics to extract and to set their extraction parameters. The general structure of the file is the following:
{\small{
\begin{tcolorbox}[rounded corners, colback=gray!30, colframe=gray!80!, title= \textbf{METRICS.in}]
\begin{itemize}[leftmargin=*]
\item[] FIRST CATEGORY NAME:
\begin{itemize}
\item[] FIRST SUBCATEGORY NAME:
\begin{itemize}
\item[] METRIC NAME 1
\item[] METRIC NAME 2
\item[] ...
\end{itemize}
\item[] SECOND SUBCATEGORY NAME:
\begin{itemize}
\item[] METRIC NAME 1
\item[] ...
\end{itemize}
\end{itemize}
\item[] SECOND CATEGORY NAME:
\begin{itemize}
    \item[] ...
\end{itemize}
\end{itemize}
\end{tcolorbox}}}
\noindent The METRICS NAME field can have \textbf{three} different input formats, depending on the corresponding definition:
\begin{enumerate}
    \item {\textbf{Metrics with no input parameters.}} The metric has to be written using the corresponding name as:
    \begin{center}
        METRIC NAME
    \end{center}
    \item {\textbf{Metrics with one input parameter.}} The metric has to be reported as:
    \begin{center}
        METRIC NAME $\rightarrow$ par1, par2, ..., parN
    \end{center}
    \item {\textbf{Metrics with two input parameters.}} The metric has to be reported as:
    \begin{center}
            METRIC NAME $\rightarrow$ [par11;par12], [par21;par22], ..., [parN1;parN2]
    \end{center}
\end{enumerate}
\begin{tcolorbox}[rounded corners, colback=brown!30, colframe=brown!80!, title= \textbf{REMEMBER}]
For cases $2$ and $3$, a given metric is computed for all the different (pairs of) input parameters. The total number of metrics computed starting from a given metric will be equal to the number of (pairs of) parameters $N$.
\end{tcolorbox}
\section{How to use it}
\subsection{Prerequisites}
The installation of TCoMX v$1.0$ requires the follows prerequisites\footnote{We refer here to the prerequisites that we were able to check.}:
\begin{itemize}
    \item Matlab{\textregistered} R2018a or newer.
    \item Windows 10 or Linux Ubuntu 20.04.1 LTS
\end{itemize}
\subsection{Installation}
Please, follow the following steps to get TCoMX ready to be used within your Matlab environment:
\begin{enumerate}
    \item Download TCoMX from \href{https://github.com/SamueleCavinato/TCoMX}{here} and extract it on your computer
    \item Move to .../TCoMX-main/input/
    \item Open the file \textit{'PATH$\_$TO$\_$TCoMX$\_$FOLDER.in}' that you find in the folder\footnote{In case your text editor does not support this type of file, you can open it using Matlab.}
    \item Substitute the line \textit{'set/your/path/to/TCoMX-main'} with the full path of the TCoMX folder on your computer
    \item Open Matlab 
    \item Add the TCoMX-main directory and subdirectories to your Matlab path. You can find instructions \href{https://it.mathworks.com/matlabcentral/answers/116177-how-to-add-a-folder-permanently-to-matlab-path}{here}.
    \item Type the command \textit{areyouready} on you matlab console and press return. If everything is properly configured, you should see the message \textit{yes, ready!}
\end{enumerate}
\subsection{Configuration}
\paragraph{\underline{CONFIGURATION OF THE INPUT PARAMETERS}}
    \begin{enumerate}
    \item Move to .../TCoMX-main/input/
    \item Open \textit{'CONFIG.in'}. Opening it with Matlab is warmly suggested since it keeps the formatting.
    \begin{enumerate}
        \item (Optional) Give a name to the execution. Type it below "Analysis ID name";
        \item Add the full path to the folder containing the DICOM RTPLAN files or the file containing the list of plans. You can use the reference dataset provided with the library. In that case, the path is the following:
        \begin{tcolorbox}
        \begin{center}
            .../TCoMX-main/database/reference$\_$dataset/\textit{TPSNAME}
        \end{center}
        \end{tcolorbox}
        where \textit{TPSNAME} needs to be raplaced either by PRECISION or RAYSTATION. The list of plans is called \textit{'list$\_$of$\_$plans.xlsx'}.
        \begin{tcolorbox}[rounded corners, colback=brown!30, colframe=brown!80!, title= \textbf{REMEMBER}]
The TCoMX library can handle three different types of input:
\begin{itemize}
    \item Folder containing all the DICOM RT-PLAN files;
    \item Folder containing a subfolder for each DICOM RT-PLAN file. 
    \item File containing the list of plans to extract the metrics from.
\end{itemize}
    \end{tcolorbox}
    \item Add the full path of the results folder where you want save the result. If it does not exist, TCoMX will create it for you
    \item Select if you want to read the list of plans from a .xlsx file (\textit{True}) or not (\textit{False}). 
    \item If you chose \textit{True} at point 1d, write the filename. 
    \end{enumerate}
    \item Close \textit{'CONFIG.in'}.
    \end{enumerate}
    \paragraph{\underline{SELECTION OF THE METRICS TO EXTRACT}}
    \begin{enumerate}
        \item Move to .../TCoMX-main/input/
        \item Open METRICS.in.
        \item You can perform the following operations:
        \begin{enumerate}
            \item Leave everything as it is and compute all the metrics;
            \item Remove some metric;
            \item  Remove a whole subcategory;
            \item Remove a whole category.
        \end{enumerate}
    \item Save and close.
    \end{enumerate}
        
    \begin{tcolorbox}[rounded corners, colback=red!30, colframe=red!80!, title= \textbf{WARNING/1}]
    The following operation are not allowed and will comprimise the correct execution of the routines in the library:
    \begin{itemize}
        \item Remove a category without removing the corresponding subcategories
        \item Remove a subcategory without removing the corresponding metrics
    \end{itemize}
    \end{tcolorbox}

      \begin{tcolorbox}[rounded corners, colback=red!30, colframe=red!80!, title= \textbf{WARNING/2}]
      \begin{itemize}
          \item In the current version, the metrics belonging to the subcategory \textit{delivery} are always computed and do not need to be added to the \textit{'METRICS.in'} file. 
          \item The computation of the TTDF requires the dose per fraction. This information is not contained in the DICOM RT-PLAN files create by Precision. In this case, you need to specify the fraction dose for each plan in a file called 'DoseFraction.xlsx' which has to be stored in the 'Dataset input folder'. An example of this file is provided together with the reference dataset. 
      \end{itemize}
    
    \end{tcolorbox}

    \begin{tcolorbox}[rounded corners, colback=ForestGreen!30, colframe=ForestGreen!80!, title= \textbf{TIPS}]
\begin{itemize}
    \item The indentation of the input file can help to visualize the content, but it is not fundamental for a correct execution.
    \item Punctuation  marks \textbf{are part of the syntax} and must be used properly. In particular:
    \begin{itemize}
        \item Each category and subcategory is followed by semicolons "$:$"
        \item No punctuation has to be but after the names of the metrics with no tunable parameters.
        \item Do not add a $\rightarrow$ after the metrics that are not endowed with tunable parameters. The library will not recognize the metric and will not extract it.
        \item Tunable parameters needs to be reported after the $\rightarrow$ and separated by a comma "$,$"
        \item Vectors of tunable parameters needs to be written after the $\rightarrow$ as $[a;b]$, and separated by commas "$,$"
    \end{itemize}

\end{itemize}
\end{tcolorbox}

  \begin{tcolorbox}[rounded corners, colback=brown!30, colframe=brown!80!, title= \textbf{REMEMBER}]
    The provided input files has been tested many times and their syntax was found to be robust over different Matlab versions and operative systems. Every time you make a change in the metrics to extract, be sure to keep the correct syntax.
    \end{tcolorbox}

\subsection{Execution}
To start the execution, type TCoMX on your Matlab console and press return. 
\chapter{Complexity metrics}
\label{ch:complexity-metrics}
The metrics included in the TCoMX library  were classified into categories and subcategories according to the features of the HT plans they are related to. Three categories have been conceived, \textit{TPS}, \textit{LOT statistics} and \textit{Sinogram}, each one containing at least one subcategory. They are all described in the following. 
\section{TPS: delivery parameters}
\subsection{Modulation Factor (MF)}
It is a dimensionless quantity computed as: 
\begin{equation}
    MF = \frac{mLOT}{maxLOT}
    \label{eq:MF}
\end{equation}
with $mLOT$ the mean LOT and $maxLOT$ the maximum LOT. These two quantities are explicitely defined in Sec.~\ref{sec:absoluteLOT}.

\subsection{Number of projection per rotation (N$_{proj, rot}$)}
Each rotation of the gantry is subdivided into N$_{proj, rot}$ descrete steps called \textit{projections}. 
\subsection{Number of projections (N$_{proj}$)}
It is the total number of projections in the treatment and corresponds to the number of rows of the sinogram stored in the RT-PLAN files minus one.
\subsection{Number of rotations (N$_{rot}$)}
It quantifies the number of gantry rotations during the treatment. It can be expressed as:
\begin{equation}
    N_{rot} = \frac{N_{proj}}{N_{proj, rot}}
\end{equation}

\subsection{Projection Time (PT)}
The duration of each projection is refered to as \textit{projection time}. It is measured in seconds (s). 
\subsection{Gantry Period (GP)}
It quantifies the duration of each rotation and it is related to PT through the following relation:
\begin{equation}
    GP =  N_{proj, rot} \times PT 
    \label{eq:GP}
\end{equation}
and it is measured in seconds (s).  
\subsection{Treatment Time (TT)}
It quantifies the duration of the treatment. It is related to the gantry period by the following relation:
\begin{equation}
    TT = GP \times N_{rot}
    \label{eq:TT}
\end{equation}
It is measured in seconds (s). 
\subsection{Field Width (FW)}
Is is defined as the distance between the Y jaws. It is measured in millimeters (mm).
\subsection{Pitch}
It is a dimensionless quantity defined as follows:
\begin{equation}
    pitch = \frac{FW}{\Delta y}
    \label{eq:pitch}
\end{equation}
where $\Delta y$ is the translation of the couch after each full gantry rotation. 
\subsection{Couch Translation (CT)}
It is defined as follows:
\begin{equation}
    CT = N_{rot}\times FW \times Pitch = N_{rot}\times \Delta y
    \label{eq:CT}
\end{equation}
It is measured in millimiters (mm): 
\subsection{Couch Speed (CS)}
It is defined as follows:
\begin{equation}
    CS = \frac{CT}{TT}
    \label{eq:CS}
\end{equation}
It is measured in millimiters per second (mm/s).
\subsection{Target Length (TL)}
It is defined as follows:
\begin{equation}
    TL = CT - FW
    \label{eq:TL}
\end{equation}
It is measured in millimiters (mm). 
\subsection{Treatment time over fraction dose (TTDF)}
It is defined as follows:
\begin{equation}
    TTDF = \frac{TT}{DF}
    \label{eq:TTDF}
\end{equation}
where DF is the dose per fraction expressed in cGy. It is measured in s/cGy.
\section{LOT statistics: absolute LOT statistics}
\subsection{LOT mean (mLOT)}
\label{sec:absoluteLOT}
It is computed as the arithmetic average of the non-zero entries of the sinogram multiplied by the projection time. The unit of measure used is the millisecond (ms).
 \subsection{LOT standard deviation (sdLOT)}
It is computed as the sample standard deviation of the non-zero entries of the sinogram multiplied by the projection time. The unit of measur used is the millisecond (ms). 
\subsection{LOT median (mdLOT)}
It is computed as the median of the non-zero entries of the sinogram multiplied by the projection time. The  unit of measure used is the millisecond (ms). 
\subsection{LOT mode (moLOT)}
It is computed as the mode of the non-zero entries of the sinogram multiplied by the projection time. The unit of measure used is the millisecond (ms). 
\subsection{LOT maximum (maxLOT)}
It is computed as the maximum of the non-zero entries of the sinogram multiplied by the projection time. The unit of measure used is the millisecond (ms). 
\subsection{LOT minimum (minLOT)}
It is computed as the minimum of the non-zero entries of the sinogram multiplied by the projection time. The  unit of measure used is the millisecond (ms). 
\subsection{LOT kurtosis (kLOT)}
It is computed as the kurtosis of the non-zero entries of the sinogram.
\subsection{LOT skewness (sLOT)}
It is computed as the skewness of the non-zero entries of the sinogram.

\subsection{Cumulative LOT Number Score (CLNS$_n$)}
It is a dimensionless metrics defined as follows:
\begin{equation}
    \label{eq:CLNSn}
    CLNS_n = \frac{1}{\sum_{ij}O_{ij}}\sum_{i=1}^{N_{proj}}\sum_{j=1}^{N_{leaves}}O^*_{ij}(n) \ \ \ \ \ \ \ \ \ \ \ \ \ \ {\begin{cases}
    O^*_{ij}(n) = 1 \ \ \ \ if \ S_{ij}\times PT < n \\
    O^*_{ij}(n) = 0 \ \ \ \ otherwise
    \end{cases}} \ \ \ \ n \in [0; PT]
\end{equation}
It counts the fractional number of LOTs smaller than $n$.

\subsection{Cumulative LOT Number Score at Projection Time (CLNS$_{pt,n}$)}
It is dimensionless metrics defined as follows:
\begin{equation}
    \label{eq:CLNSptn}
    CLNS_{pt, n} = \frac{1}{\sum_{ij}O_{ij}}\sum_{i=1}^{N_{proj}}\sum_{j=1}^{N_{leaves}}O^*_{ij}(n) \ \ \ \ \ \ \  {\begin{cases}
    O^{(pt*)}_{ij}(n) = 1 \ \ \ \ if \ S_{ij}\times PT > PT - n \\
    O^{(pt*)}_{ij}(n) = 0 \ \ \ \ otherwise
    \end{cases}} \ \ \ \ n \in [0, PT]
\end{equation}
which counts the fractional number of LOTs which are closer than $n$ ms to the projection time. 

\section{LOT statistics: relative LOT statistics}
\subsection{FLOT mean (mFLOT)}
It is computed as the arithmetic average of the non-zero entries of the sinogram. It is a dimensionless metric.
 \subsection{FLOT standard deviation (sdFLOT)}
It is computed as the sample standard deviation of the non-zero entries of the sinogram. It is a dimensionless metric.
\subsection{FLOT median (mdFLOT)}
It is computed as the median of the non-zero entries of the sinogram. It is a dimensionless metric.
\subsection{FLOT mode (moFLOT)}
It is computed as the mode of the non-zero entries of the sinogram. It is a dimensionless metric.
\subsection{FLOT maximum (maxFLOT)}
It is computed as the maximum of the non-zero entries of the sinogram. It is a dimensionless metric.
\subsection{FLOT minimum (minFLOT)}
It is computed as the minimum of the non-zero entries of the sinogram. It is a dimensionless metric.

\subsection{Cumulative FLOT Number Score (CFNS$_n$)}
It is a dimensionless metric defined as follows:
\begin{equation}
    \label{eq:CFNS}
    CFNS_n = \frac{1}{\sum_{ij}O_{ij}}\sum_{i=1}^{N_{proj}}\sum_{j=1}^{N_{leaves}}O^*_{ij}(n) \ \ \ \ \ \ \ \ \ \ \ \ \ \ {\begin{cases}
    O^*_{ij}(n) = 1 \ \ \ \ if \ S_{ij} < n \\
    O^*_{ij}(n) = 0 \ \ \ \ otherwise
    \end{cases}} \ \ \ \ n \in [0;1]
\end{equation}
It counts the fraction of FLOTs smaller than $n$.

\section{Sinogram: geometry}
\subsection{Leaves with n Open Nearest Neighbors (LnNS)}
It was introduced by Santos et al. in~\cite{santos:complexity-metrics} and it is a dimensionless metric defined as follows:
\begin{equation}
    LnNS =\frac{1}{N_{proj}}\sum_{i=1}^{N_{proj}}\left[ \frac{N_O(n)}{N_O}\right]_i \times 100 \%
\end{equation}
where $N_{O}(n)$ indicates the number of open leaves with $n$ open neighbors and $N_O$ the number of open leaves. The sub-index $i$ indicates that the ratio has to be evaluated at each projection. The plan value is obtained as the average over all the projections. The index $n$ can take three different values: $0$, $1$ and $2$. \\


\subsection{Treatment Area (TA)}
It is defined as follows:
\begin{equation}
    TA = \frac{1}{N_{proj}}\sum_{i=1}^{N_{proj}} \left(
    |R_i - L_i|+1\right)
\end{equation}
where $R_i$ and $L_i$ are the right-most and left-most open leaves at projection $i$, respectively.The plan value is obtained by averaging over all the projections. Since the LOTs values are not considered in this definition, TA represents the average cumulative open area. It is measured in number-of-leaves.

\subsection{Number of Connected Components (nCC)}
\begin{figure}[t!]
    \centering
    \includegraphics[width=0.75\textwidth]{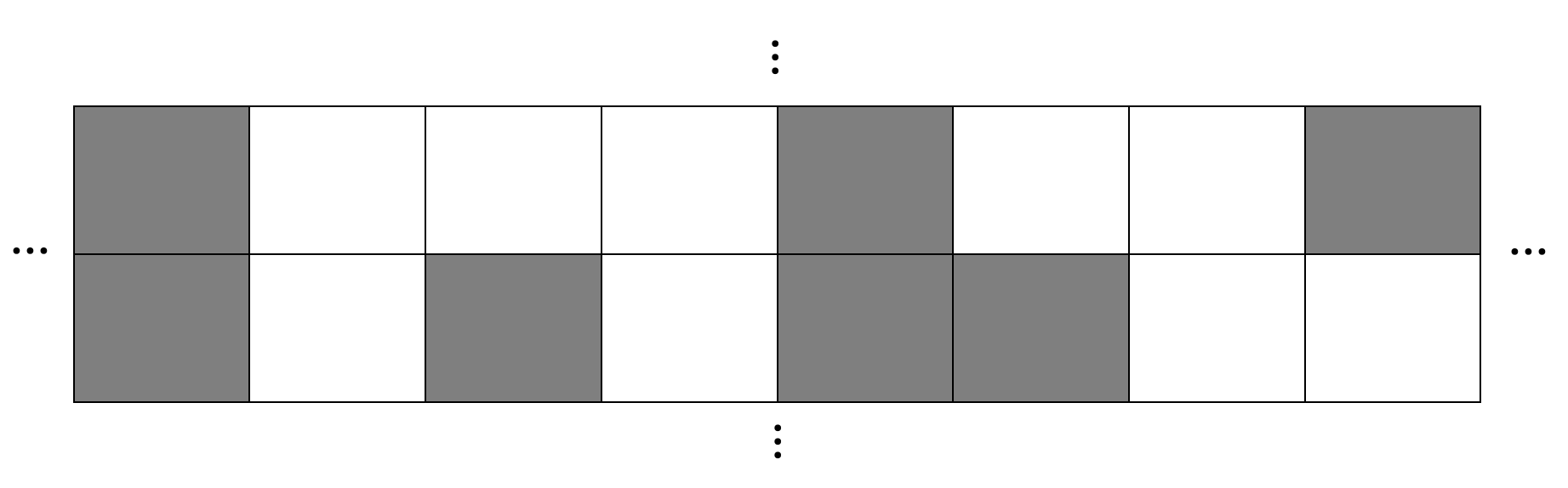}
    \caption{Schematic representation of two projections with different numbers of connected components. White and gray squares represent open and closed leaves, respectively. The upper projection has 2 connected components, the lower one has 3.
}
    \label{fig:nCC}
\end{figure}

It is a dimensionless metric which counts the number of independent groups of connected leaves inside the treatment area. It is computed at each projection and then averaged over all the projections to get the plan value. In Fig.~\ref{fig:nCC} an example of two projections with different numbers of connected components is shown.

\subsection{Length of the Connected Components (lengthCC)}
It is defined as follows:
\begin{equation}
    \label{eq:lengthCC}
    lengthCC = \frac{1}{\sum_{i=1}^{N_{proj}}nCC_i}\sum_{i=1}^{N_{proj}} \sum_{k=1}^{nCC_i} |L_k - R_k|+1 
\end{equation}
where $L_k$ and $R_k$  are the positions of the leftmost and rightmost open leaves of the $k-th$ connected component at the $i-th$ projection, respectively. It is measured in number-of-leaves. The plan value measures the average length of the connected components. \\    
In Fig.~\ref{fig:nCC}, four examples of connected components (white blocks) with different lengths are shown. The two connected components of the upper projection have length 3 and 2, respectively. The three connected components of the lower projection have length 1, 1 and 2, respectively. The plan value is obtained by averaging over the total number of connected components of the sinogram.

\subsection{Fraction of Discontinuous Projections (fDISC)}
It is a dimensionless metric defined as follows:
\begin{equation}
    \label{eq:fDISC}
    fDISC = \frac{1}{N_{proj}}\sum_{i=1}^{N_{proj}}[nCC_i>1]
\end{equation}
where the term $[nCC_i >1]$ is $0$ if $nCC_i = 0$ or $1$, $1$ otherwise. It counts the fraction of projections with two or more connected components, namely all the projections having at least one closed leaf within the treatment area.

\subsection{Closed Leaf Score (CLS)}
It was introduced in \cite{santos:complexity-metrics} and it is a dimensionless defined as follows:
\begin{equation}
    CLS = \frac{1}{N_{proj}}\sum_{i=1}^{N_{proj}}\left[\frac{N_{leaves} - \sum_{j}^{N_{leaves}} O_{ij}}{N_{leaves}}\right]
\end{equation}
It can take values in $[0,1]$, being $1$ when all leaves remain closed. 

\subsection{Closed Leaf Score within the treatment area $(CLS_{in})$}
It was introduced in \cite{santos:complexity-metrics} with this definition:
\begin{equation}
\label{eq:CLSin}
    CLS_{in} = \frac{1}{N_{proj}}\sum_{i=1}^{N_{proj}}\left[\frac{TA_i - \sum_{j}^{N_{leaves}} O_{ij}}{N_{leaves}}\right]
\end{equation}
A more general form has been devised and three variants have been introduced:
\begin{enumerate}
    \item $CLS_{in,area}$ : the number of closed leaves within the treatment area at projecton $i$ is normalized by $TA_i$ instead of $N_{leaves}$;
    \item $CLS_{in,disc}$ : it is computed by considering the discontinuous projections $(nCC_i>1)$ only, namely the first summation in Eq.~(\ref{eq:CLSin}) runs from 1 to the total number of discontinuous projections and is then divided by the number of discontinuous projections instead of by $N_{proj}$ ;
    \item $CLS_{in,area, disc}$ : it corresponds to the combination of the $CLS_{in,area}$ with the $CLS_{in,disc}$; 

\end{enumerate}
The three quantities are in general correlated with each other. However, the different definitions should help to characterize the geometry of the leaf openings in a more intuitive way. 

\subsection{Centroid}
It is defined as follows:
\begin{equation}
    \label{eq:centroid}
    centroid = \frac{1}{N_{proj}} \sum_{i=1}^{N_{proj}}\frac{1}{\sum_{j=1}^{N_{leaves}}O_{ij}} \left[ \sum_{j=1}^{N_{leaves}} O_{ij} P_j\right]
\end{equation}
The centroid is measured in number-of-leaves. The plan value is obtained by averaging over $N_{proj}$. The centroid of a plan represents the mean of the average positions of the open leaves at each projection.

\section{Sinogram: modulation}
\subsection{Leaf Open Time Variability (LOTV)}
It was introduced in~\cite{santos:complexity-metrics} and it is a dimensionless metric defined as follows:
\begin{equation}
    LOTV = \frac{1}{N_{leaves}}\sum_{j=1}^{N_{leaves}}\left[\frac{\sum_{i=1}^{N_{proj}-1} max(S_{j})- |S_{ij}-S_{i+1,j}|}{(N_{proj}-1)\times max(S_{j})}\right]
\end{equation}
where $S_j$ marks the j-th column (leaf) of the sinogram. It takes values in $[0,1]$, being $1$ when all the leaves have the same opening time at each projection. For the leaves which do not open during the treatment it is set to $1$ by definition.
\subsection{Extended Leaf Open Time Variability (ELOTV$_{\Delta p}$)}
It is a dimensionless metric defined as follows: 
\begin{equation}
    ELOTV_{\Delta p}= \frac{1}{N_{leaves}}\sum_{j=1}^{N_{leaves}}\left[\frac{\sum_{i=1}^{N_{proj}-{\Delta p}} |S_{ij}-S_{i+{\Delta p},j}|}{(N_{proj} -\Delta p) \times max(S_{j})}\right]
\end{equation}
where $\Delta p$ is the projection step and $S_j$ the j-th column (leaf) of the sinogram. It takes values in $[0,1]$, being $0$ when all the leaves have the same opening time at each projection. The $ELOTV_{\Delta p}$ is first evaluated for each leaf and then averaged over all the leaves to obtain the plan value. For the leaves which do not open during the treatment it is set to $0$ by definition. It is worth noticing that the $ELOTV_{\Delta p}$ includes the $LOTV$ as a special case. In particular:
\begin{equation}
    LOTV = 1 - ELOTV_1
\end{equation}
Compared to the $LOTV$ it shows two main differences:
\begin{itemize}
    \item It is positively correlated with the inter-projection (F)LOT variability;
    \item It allows the comparison of projections which lie at arbitrary distances.
\end{itemize}

\subsection{Plan Sinogram Time Variation (PSTV)}
It was introduced by Santos et al. in~\cite{santos:complexity-metrics} and it is a dimensionless metrics defined as follows:
\begin{equation}
    \label{eq:PSTV}
    PSTV = \frac{1}{N_{proj}-1}\sum_{i=1}^{N_{proj}-1} \sum_{j=1}^{N_{leaves}-1} |S_{i+1,j} - S_{i,j}| + |S_{i,j+1} - S_{i,j}|
\end{equation}
 which is the arithmetic average over all the projections of the sum of FLOTs difference between pairs of adjacent leaves and projections.
Higher values of PSTV value correspond to a higher variability of the (F)LOT across the leaves and the projections.

\subsection{Extended Plan Sinogram Time Variation (EPSTV$_{\Delta p, \Delta l}$)}
It is a dimensionless metric defined as follows:
\begin{equation}
    {EPSTV_{\Delta p, \Delta l}} = \frac{1}{N_{proj}-\Delta p}\sum_{i=1}^{N_{proj}-\Delta p} \sum_{j=1}^{N_{leaves}-\Delta l} |S_{i+\Delta p,j} - S_{i,j}| + |S_{i,j+\Delta l} - S_{i,j}|
\end{equation}
The $EPSTV_{\Delta p, \Delta l}$ contains the $PSTV$ defined in Eq.~(\ref{eq:PSTV}) as a special case, namely: 
\begin{equation}
    PSTV = EPSTV_{1,1}
\end{equation}
This new formulation allows the comparison of projections/leaves which lie at arbitrary distances. Furthermore, there are two special cases included in it, namely:
\begin{itemize}
    \item \textit{EPSTV$_{0, \Delta l}$}: time variation along the leaves direction;
    \item \textit{EPSTV$_{\Delta p, 0}$}: time variation along the projection direction.
\end{itemize}
In general, a higher $EPSTV_{\Delta p, \Delta l}$ corresponds to higher inter-leaf and/or inter-projection (F)LOT variability. 

\subsection{Modulation Index (MI) }
The \textit{Modulation Index} was introduced by Park and collegues~\cite{Park_2014_MI} for VMAT plans. It was adapted to HT plans by Santos and colleques~\cite{santos:complexity-metrics}. It is a dimensionless metric defined as follows: 
\begin{equation}
    \label{eq:MI}
    MI = \int_0^{2\sigma}Z(f)df
\end{equation}
where Z(f) is defined as:
\begin{equation}
    \label{eq:Z(f)}
    Z(f) = \frac{Z_x(f) + Z_y(f) + Z_{xy}(f) + Z_{yx}(f)}{4}
\end{equation}
with each $Z_k(f)$ defined as:
\begin{equation}
    \label{eq:Z_k(f)}
    Z_k(f) = \frac{1}{N_{proj}}N_k(f; \Delta t_k > f\sigma)
\end{equation}
where $k = \{x,y,xy, yx\}$ represents four different directions of the sinogram $S$ (leaves, projections, diagonal and anti-diagonal), $\Delta t_k$ represents the time variation between adjacent elements in the $k$ direction, $f$ a fraction of the  $sdFLOT$ and $N_k(f, \Delta t_k > f\sigma)$ the number of FLOTs exeeding this fraction. 
Higher values of $MI$ should correspond to a higher plan modulation. 
\subsection{Number of Openings and Closures (nOC)}
It is a dimensionless metric and it is computed by counting the number of times each leaf opens and closes during the treatment. The plan value is obtained by averaging over the leaves and normalizing by the total number of projections. 
The number of openings and closures is computed considering that each (F)LOT is centered with respect to the projection time. 
\begin{figure}[ht!]
    \centering
    \includegraphics[width=0.75\textwidth]{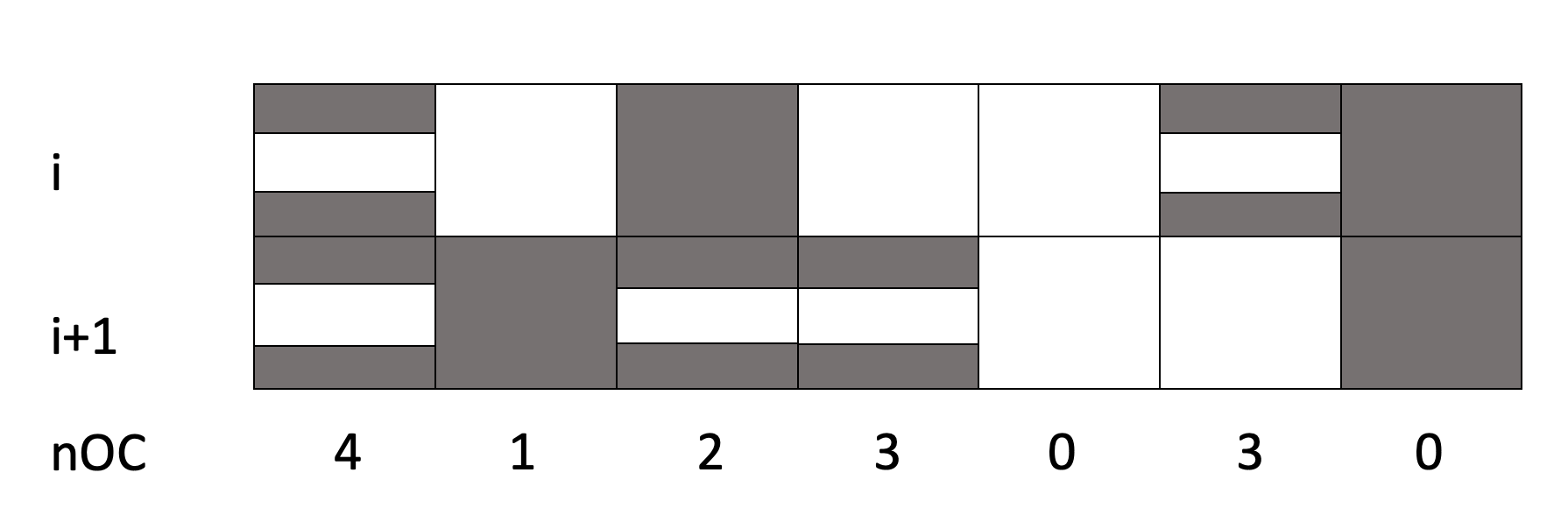}
    \caption{Schematic representation of possible combinations of the LOTs leading to different numbers of openings and closures. Rows represent two adjacent projections, while columns show seven possible leaf movements. White and gray spaces represent open and closed leaves, respectively. Each LOT is centered about the middle point of the corresponding projection. The number under each leaf represents the cumulative number of openings and closures for each leaf over the two projections.}
    \label{fig:nOC}
\end{figure}
In Fig.~\ref{fig:nOC}, a schematic of the different conditions that might be encountered during the treatment is shown. 
This metric is associated with the mechanical stress of the MLC during the treatment. Moreover, this metric is related to the CLS (average fraction of closed leaves per projectoin) by the following approximate linear relationship:
\begin{equation}
    CLS \approx 1-0.5\times nOC
\end{equation}
This is due to the fact that fully closed ($S_{ij} = 0$) have $nOC_{ij} = 0$ and fully open ($S_{ij}=1$) appear only in a negligible amount. Therefore, each $0<S_{ij}<1$ correspond to $nOC_{ij}=2$.

\begin{figure}[ht!]
    \centering
    \includegraphics[width=0.3\textwidth, height=15cm]{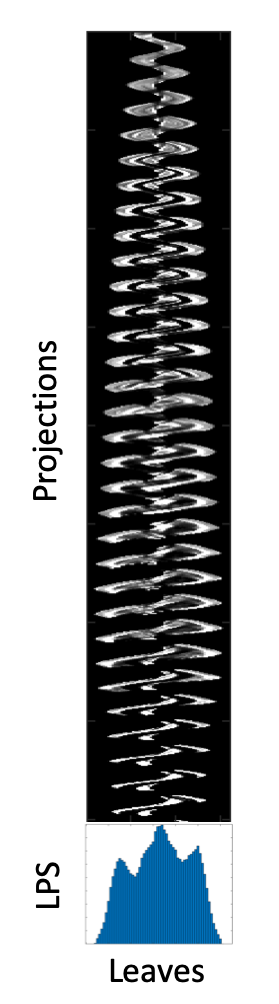}
    \caption{(Upper panel) Example sinogram. Rows correspond to the projections, columns to the leaves. (Lower Panel) Example of LPS. Each bar corresponds to a leaf (column of the sinogram), the height corresponds to the sum over all the projections properly normalized by $N_{proj}$.}
    \label{fig:LPS}
\end{figure}

\subsection{Mean Sinogram Asymmetry (mSA)}
It is defined as follows:
\begin{equation}
    \label{eq:mLPS}
    MSA =  \frac{\sum_{j=1}^{N_{leaves}} P_j\times LPS_j}{\sum_{j=1}^{N_{leaves}} LPS_j}
\end{equation}
where $LPS_j$ is called \textit{Leaf Projected Sinogram} and is defined as:
\begin{equation}
    \label{eq:LPS_j}
    LPS_j = \sum_{i=1}^{N_{proj}}\frac{S_{ij}}{N_{proj}}
\end{equation}
An example of $LPS$ is shown in Fig.~\ref{fig:LPS}. The MSA represents the weighted average displacement of the (F)LOTs from the vertical axis passing through the center of the sinogram and it is measured in number-of-leaves. It allows to highlight asymmetries of the sinogram. 

\subsection{Sinogram intensity (SI)}
Starting from the LPS defined in Eq.~(\ref{eq:LPS_j}), it is possible to compute another metric related to it, namely the \textit{Sinogram Intensity}. In particular, the mean sinogram intensity (mSI), is computed as the average of the LPS values over the MLC leaves. Leaves with $LPS = 0$ are included in the computation and this fact distinguishes mSI from mLOT (mFLOT). In addition to mSI, also its standard deviation (sdSI) and the median (mdSI) are computed. 



\chapter{Reference dataset}
\label{ch:reference_dataset}

\section{Description}
The TCoMX library is provided with a dataset that can be freely used by the user to test the functioning. It contains 18 DICOM RT-PLAN files corresponding to 9 patients. Nine plans where realized using Precision TPS, the other 9 using RayStation. All the plans have been anonymized. Plans corresponding to each TPS where renamed using a sequential numbering, namely $RP_{0n}$, with $n = [1,9]$ to keep the correspondence between plans of the same patient. \\
The plans correspond to three different treatments: Prostate, Head and neck and Meshothelioma. A total amount of 66 complexity metrics were extracted. A summary of the metrics extracted is reported in Tab.~\ref{tab:metrics_extracted}. The results are reported in Sec~\ref{sec:results}.

\begin{table}[h]
\centering
\arrayrulecolor{black}
\begin{tabular}{l|l|l} 
\hline
\textbf{CATEGORY} & \textbf{SUBGATEGORY} & \textbf{METRIC NAME} \\ 
\hline
\textbf{TPS} & \textbf{Delivery} & \begin{tabular}[c]{@{}l@{}}Pitch, FW, PT, GP, TT, TL, CS, CT, N$_{proj}$,\\ N$_{rot}$, N$_{rot, proj}$ MF, TTDF\end{tabular} \\ 
\hline
\multirow{2}{*}{\begin{tabular}[c]{@{}l@{}}\textbf{LOT} \\\textbf{STATISTICS}\end{tabular}} & \textbf{Absolute LOT} & \begin{tabular}[c]{@{}l@{}}mLOT, sdLOT, mdLOT, moLOT,  kLOT, \\ sLOT, minLOT, maxLOT, CLNS$_{100}$,CLNS$_{50}$,\\ CLNS$_{30}$, CLNS$_{20}$, CLNS$_{pt,20}$ \end{tabular} \\ 
\arrayrulecolor{black}\cline{2-3}
 & \textbf{Relative LOT} & \begin{tabular}[c]{@{}l@{}}mFLOT, sdFLOT, mdFLOT, moFLOT, \\ minFLOT,maxFLOT, CFNS$_{5}$, CFNS$_{10}$, CFNS$_{25}$,\\ CFNS$_{50}$, CFNS$_{75}$, CFNS$_{90}$, CFNS$_{95}$ \end{tabular} \\ 
\arrayrulecolor{black}\hline
\multirow{2}{*}{\textbf{SINOGRAM}} & \textbf{Geometry} & \begin{tabular}[c]{@{}l@{}}L$0$NS, L$1$NS, L$2$NS, nCC, lengthCC, TA, fDISC, \\ CLS, CLS$_{in}$, CLS$_{in,disc}$,  CLS$_{in,area}$, CLS$_{in,area,disc}$, \\ centroid\end{tabular} \\ 
\arrayrulecolor{black}\cline{2-3}
 & \textbf{Modulation} & \begin{tabular}[c]{@{}l@{}}nOC, PSTV, EPSTV$_{1,1}$, EPSTV$_{0,1}$, EPSTV$_{1,0}$,\\LOTV, ELOTV$_{1}$, ELOTV$_{3}$, ELOTV$_{5}$, \\MI, mSI, mdSI, sdSI, MSA\end{tabular} \\
\arrayrulecolor{black}\hline
\end{tabular}
\caption{List of the metrics extracted from the reference dataset.}
\label{tab:metrics_extracted}
\end{table}

\newpage
\section{Results}
\label{sec:results}

\begin{table}[h]
\centering
\arrayrulecolor{black}
\begin{tabular}{l|rrr|rrr|rrr|l} 
\toprule
\multicolumn{1}{c}{} & \multicolumn{3}{c}{\textbf{Prostate}} & \multicolumn{3}{c}{\textbf{Head and neck}} & \multicolumn{3}{c}{\textbf{Mesothelioma}} & \multicolumn{1}{c}{} \\
\cline{1-10}\arrayrulecolor{black}\cline{11-11}
\multicolumn{1}{l}{} & \multicolumn{1}{l}{\textbf{RP01}} & \multicolumn{1}{l}{\textbf{RP02}} & \multicolumn{1}{l}{\textbf{RP03}} & \multicolumn{1}{l}{\textbf{RP04}} & \multicolumn{1}{l}{\textbf{RP05}} & \multicolumn{1}{l}{\textbf{RP06}} & \multicolumn{1}{l}{\textbf{RP07}} & \multicolumn{1}{l}{\textbf{RP08}} & \multicolumn{1}{l}{\textbf{RP09}} &  \\ 
\cline{1-10}\arrayrulecolor{black}\cline{11-11}
\textbf{Pitch} & 0.43 & 0.43 & 0.43 & 0.45 & 0.43 & 0.44 & 0.42 & 0.40 & 0.33 & \multicolumn{1}{l}{\multirow{13}{*}{\textbf{\rotatebox{-90}{PRECISION}}}} \\
\textbf{FW} & 25.12 & 25.12 & 25.12 & 25.12 & 25.12 & 25.12 & 25.12 & 25.12 & 25.12 & \multicolumn{1}{l}{} \\
\textbf{PT} & 0.62 & 0.56 & 0.51 & 0.46 & 0.30 & 0.28 & 0.41 & 0.32 & 0.28 & \multicolumn{1}{l}{} \\ 
\textbf{GP} & 31.40 & 28.40 & 25.90 & 23.70 & 15.20 & 14.20 & 20.90 & 16.10 & 14.50 & \multicolumn{1}{l}{} \\
\textbf{TT} & 216.72 & 261.73 & 227.51 & 336.91 & 306.98 & 175.69 & 658.55 & 482.37 & 483.05 & \multicolumn{1}{l}{} \\
\textbf{TL} & 49.43 & 74.42 & 69.76 & 135.57 & 193.03 & 110.39 & 305.74 & 275.93 & 253.55 & \multicolumn{1}{l}{} \\
\textbf{CS} & 0.34 & 0.38 & 0.42 & 0.48 & 0.71 & 0.77 & 0.50 & 0.62 & 0.58 & \multicolumn{1}{l}{} \\
\textbf{CT} & 74.55 & 99.54 & 94.88 & 160.69 & 218.15 & 135.51 & 330.86 & 301.05 & 278.67 & \multicolumn{1}{l}{} \\
\textbf{N$_{proj}$} & 352 & 470 & 448 & 725 & 1030 & 631 & 1607 & 1528 & 1699 & \multicolumn{1}{l}{} \\
\textbf{N$_{rot}$} & 6.90 & 9.22 & 8.78 & 14.22 & 20.20 & 12.37 & 31.51 & 29.96 & 33.31 & \multicolumn{1}{l}{} \\
\textbf{N$_{proj,rot}$} & 51 & 51 & 51 & 51 & 51 & 51 & 51 & 51 & 51 & \multicolumn{1}{l}{} \\
\textbf{MF} & 1.88 & 1.85 & 1.77 & 1.89 & 1.78 & 1.48 & 1.79 & 1.44 & 1.60 & \multicolumn{1}{l}{} \\
\textbf{TTDF} & 1.08 & 1.23 & 0.87 & 1.59 & 1.38 & 1.66 & 3.29 & 7.99 & 2.68 & \multicolumn{1}{l}{} \\ 
\arrayrulecolor{black}\cline{1-10}\arrayrulecolor{black}\cline{11-11}
\multicolumn{1}{l}{} & \multicolumn{1}{l}{} & \multicolumn{1}{l}{} & \multicolumn{1}{l}{} & \multicolumn{1}{l}{} & \multicolumn{1}{l}{} & \multicolumn{1}{l}{} & \multicolumn{1}{l}{} & \multicolumn{1}{l}{} & \multicolumn{1}{l}{} &  \\ 
\arrayrulecolor{black}\hline
\textbf{Pitch} & 0.43 & 0.43 & 0.43 & 0.43 & 0.43 & 0.43 & 0.43 & 0.43 & 0.43 & \multirow{13}{*}{\textbf{\rotatebox{-90}{RAYSTATION}}} \\
\textbf{FW} & 25.12 & 25.12 & 25.12 & 25.12 & 25.12 & 25.12 & 25.12 & 25.12 & 25.12 &  \\
\textbf{PT} & 0.84 & 0.51 & 0.42 & 0.34 & 0.37 & 0.35 & 0.31 & 0.37 & 0.37 &  \\
\textbf{GP} & 42.80 & 25.80 & 21.40 & 17.40 & 18.90 & 17.70 & 15.90 & 19.00 & 18.80 &  \\
\textbf{TT} & 337.36 & 263.06 & 199.73 & 274.65 & 399.49 & 240.86 & 498.82 & 547.27 & 498.38 &  \\
\textbf{TL} & 60.02 & 85.01 & 75.69 & 145.38 & 203.20 & 121.87 & 313.75 & 286.01 & 261.23 &  \\
\textbf{CS} & 0.25 & 0.42 & 0.50 & 0.62 & 0.57 & 0.61 & 0.68 & 0.57 & 0.57 &  \\
\textbf{CT} & 85.14 & 110.13 & 100.81 & 170.50 & 228.32 & 146.99 & 338.87 & 311.13 & 286.35 &  \\
\textbf{N$_{proj}$} & 402 & 520 & 476 & 805 & 1078 & 694 & 1600 & 1469 & 1352 &  \\
\textbf{N$_{rot}$} & 7.88 & 10.20 & 9.33 & 15.78 & 21.14 & 13.61 & 31.37 & 28.80 & 26.51 &  \\
\textbf{N$_{proj,rot}$} & 51 & 51 & 51 & 51 & 51 & 51 & 51 & 51 & 51 &  \\
\textbf{MF} & 2.42 & 1.91 & 1.54 & 1.83 & 2.04 & 2.33 & 1.54 & 1.66 & 1.40 &  \\
\textbf{TTDF} & 1.30 & 1.06 & 0.81 & 2.01 & 2.11 & 2.08 & 3.79 & 12.17 & 9.15 &  \\
\bottomrule
\end{tabular}
\caption{Delivery parameters extracted from the plans in the reference dataset.}
\end{table}

\begin{table}[h!]
\begin{tabular}{l|rrr|rrr|rrr|l} 
\hline
\multicolumn{1}{c}{} & \multicolumn{3}{c}{\textbf{Prostate}} & \multicolumn{3}{c}{\textbf{Head and neck}} & \multicolumn{3}{c}{\textbf{Mesothelioma}} & \multicolumn{1}{c}{} \\
\cline{1-10}\arrayrulecolor{black}\cline{11-11}
\multicolumn{1}{l}{} & \multicolumn{1}{l}{\textbf{RP01}} & \multicolumn{1}{l}{\textbf{RP02}} & \multicolumn{1}{l}{\textbf{RP03}} & \multicolumn{1}{l}{\textbf{RP04}} & \multicolumn{1}{l}{\textbf{RP05}} & \multicolumn{1}{l}{\textbf{RP06}} & \multicolumn{1}{l}{\textbf{RP07}} & \multicolumn{1}{l}{\textbf{RP08}} & \multicolumn{1}{l}{\textbf{RP09}} &  \\ 
\hline
\textbf{mLOT} & 327.33 & 300.27 & 286.00 & 246.04 & 167.44 & 186.49 & 228.35 & 217.80 & 177.68 & \multirow{13}{*}{\textbf{\rotatebox{-90}{PRECISION}}} \\
\textbf{sdLOT} & 150.58 & 152.01 & 130.19 & 109.90 & 80.42 & 92.79 & 133.47 & 105.01 & 100.52 &  \\
\textbf{mdLOT} & 345.83 & 297.06 & 291.83 & 231.10 & 168.81 & 211.54 & 231.48 & 251.66 & 188.49 &  \\
\textbf{moLOT} & 615.14 & 555.74 & 506.10 & 464.43 & 297.91 & 276.67 & 408.92 & 314.56 & 283.50 &  \\
\textbf{kLOT} & 2.57 & 2.29 & 2.44 & 2.83 & 2.01 & 1.64 & 1.66 & 1.81 & 1.41 &  \\
\textbf{sLOT} & -0.16 & 0.08 & -0.10 & 0.49 & 0.03 & -0.44 & -0.02 & -0.58 & -0.22 &  \\
\textbf{minLOT} & 18.05 & 18.04 & 18.03 & 18.31 & 18.03 & 18.02 & 18.00 & 18.00 & 18.00 &  \\
\textbf{maxLOT} & 615.14 & 555.74 & 506.10 & 464.43 & 297.91 & 276.67 & 408.92 & 314.56 & 283.50 &  \\
\textbf{CLNS$_{100}$} & 0.10 & 0.13 & 0.11 & 0.08 & 0.24 & 0.25 & 0.24 & 0.21 & 0.32 &  \\
\textbf{CLNS$_{50}$} & 0.05 & 0.06 & 0.05 & 0.03 & 0.08 & 0.12 & 0.12 & 0.10 & 0.16 &  \\
\textbf{CLNS$_{30}$} & 0.02 & 0.02 & 0.02 & 0.01 & 0.03 & 0.05 & 0.06 & 0.05 & 0.07 &  \\
\textbf{CLNS$_{20}$} & 0.00 & 0.00 & 0.00 & 0.00 & 0.01 & 0.01 & 0.01 & 0.01 & 0.01 &  \\
\textbf{CLNS$_{pt,20}$} & 0.07 & 0.13 & 0.11 & 0.11 & 0.13 & 0.42 & 0.22 & 0.44 & 0.39 &  \\ 
\hline
\multicolumn{1}{l}{} & \multicolumn{1}{l}{} & \multicolumn{1}{l}{} & \multicolumn{1}{l}{} & \multicolumn{1}{l}{} & \multicolumn{1}{l}{} & \multicolumn{1}{l}{} & \multicolumn{1}{l}{} & \multicolumn{1}{l}{} & \multicolumn{1}{l}{} &  \\ 
\hline
\textbf{mLOT} & 347.19 & 265.31 & 271.70 & 186.18 & 181.78 & 148.71 & 202.22 & 224.57 & 264.12 & \multirow{13}{*}{\textbf{\rotatebox{-90}{RAYSTATION}}} \\
\textbf{sdLOT} & 144.30 & 112.35 & 119.07 & 95.20 & 137.34 & 62.93 & 115.99 & 138.20 & 132.93 &  \\
\textbf{mdLOT} & 347.00 & 264.77 & 264.30 & 177.45 & 81.77 & 134.55 & 311.76 & 213.21 & 368.63 &  \\
\textbf{moLOT} & 839.22 & 505.88 & 419.61 & 341.18 & 370.59 & 347.06 & 311.76 & 372.55 & 368.63 &  \\
\textbf{kLOT} & 4.33 & 2.96 & 1.85 & 1.99 & 1.41 & 4.97 & 1.11 & 1.18 & 1.49 &  \\
\textbf{sLOT} & 0.44 & 0.39 & -0.14 & 0.40 & 0.51 & 1.39 & -0.18 & -0.02 & -0.60 &  \\
\textbf{minLOT} & 66.42 & 68.50 & 64.25 & 63.08 & 63.03 & 74.77 & 63.07 & 63.03 & 63.06 &  \\
\textbf{maxLOT} & 839.22 & 505.88 & 419.61 & 341.18 & 370.59 & 347.06 & 311.76 & 372.55 & 368.63 &  \\
\textbf{CLNS$_{100}$} & 0.08 & 0.09 & 0.13 & 0.25 & 0.53 & 0.27 & 0.39 & 0.36 & 0.25 &  \\
\textbf{CLNS$_{50}$} & 0.00 & 0.00 & 0.00 & 0.00 & 0.00 & 0.00 & 0.00 & 0.00 & 0.00 &  \\
\textbf{CLNS$_{30}$} & 0.00 & 0.00 & 0.00 & 0.00 & 0.00 & 0.00 & 0.00 & 0.00 & 0.00 &  \\
\textbf{CLNS$_{20}$} & 0.00 & 0.00 & 0.00 & 0.00 & 0.00 & 0.00 & 0.00 & 0.00 & 0.00 &  \\
\textbf{CLNS$_{pt,20}$} & 0.01 & 0.09 & 0.31 & 0.20 & 0.31 & 0.04 & 0.52 & 0.43 & 0.59 &  \\
\hline
\end{tabular}
\caption{Absolute LOT statistics extracted from the plans in the reference dataset.}
\end{table}

\begin{table}
\centering
\begin{tabular}{l|rrr|rrr|rrr|l} 
\hline
\multicolumn{1}{c}{} & \multicolumn{3}{c}{\textbf{Prostate}} & \multicolumn{3}{c}{\textbf{Head and neck}} & \multicolumn{3}{c}{\textbf{Mesothelioma}} & \multicolumn{1}{c}{} \\
\cline{1-10}\arrayrulecolor{black}\cline{11-11}
\multicolumn{1}{l}{} & \multicolumn{1}{l}{\textbf{RP01}} & \multicolumn{1}{l}{\textbf{RP02}} & \multicolumn{1}{l}{\textbf{RP03}} & \multicolumn{1}{l}{\textbf{RP04}} & \multicolumn{1}{l}{\textbf{RP05}} & \multicolumn{1}{l}{\textbf{RP06}} & \multicolumn{1}{l}{\textbf{RP07}} & \multicolumn{1}{l}{\textbf{RP08}} & \multicolumn{1}{l}{\textbf{RP09}} &  \\ 
\hline
\textbf{mFLOT} & 0.53 & 0.54 & 0.56 & 0.53 & 0.56 & 0.67 & 0.56 & 0.69 & 0.62 & \multirow{13}{*}{\textbf{\rotatebox{-90}{PRECISION}}} \\
\textbf{sdFLOT} & 0.24 & 0.27 & 0.26 & 0.24 & 0.27 & 0.33 & 0.33 & 0.33 & 0.35 &  \\
\textbf{mdFLOT} & 0.56 & 0.53 & 0.57 & 0.50 & 0.57 & 0.76 & 0.56 & 0.80 & 0.66 &  \\
\textbf{moFLOT} & 1.00 & 1.00 & 1.00 & 1.00 & 1.00 & 0.99 & 1.00 & 1.00 & 1.00 &  \\
\textbf{minFLOT} & 0.03 & 0.03 & 0.04 & 0.04 & 0.06 & 0.06 & 0.04 & 0.06 & 0.06 &  \\
\textbf{maxFLOT} & 1.00 & 1.00 & 1.00 & 1.00 & 1.00 & 0.99 & 1.00 & 1.00 & 1.00 &  \\
\textbf{CFNS$_{5}$} & 0.02 & 0.02 & 0.01 & 0.00 & 0.00 & 0.00 & 0.01 & 0.00 & 0.00 &  \\
\textbf{CFNS$_{10}$} & 0.06 & 0.07 & 0.05 & 0.02 & 0.03 & 0.04 & 0.10 & 0.05 & 0.06 &  \\
\textbf{CFNS$_{25}$} & 0.17 & 0.17 & 0.14 & 0.11 & 0.16 & 0.18 & 0.25 & 0.17 & 0.23 &  \\
\textbf{CFNS$_{50}$} & 0.37 & 0.43 & 0.36 & 0.51 & 0.42 & 0.35 & 0.44 & 0.31 & 0.41 &  \\
\textbf{CFNS$_{75}$} & 0.85 & 0.79 & 0.78 & 0.83 & 0.74 & 0.50 & 0.68 & 0.47 & 0.54 &  \\
\textbf{CFNS$_{90}$} & 0.92 & 0.85 & 0.87 & 0.88 & 0.85 & 0.57 & 0.76 & 0.55 & 0.60 &  \\
\textbf{CFNS$_{95}$} & 0.93 & 0.86 & 0.88 & 0.89 & 0.87 & 0.59 & 0.78 & 0.57 & 0.61 &  \\ 
\hline
\multicolumn{1}{l}{} & \multicolumn{1}{l}{} & \multicolumn{1}{l}{} & \multicolumn{1}{l}{} & \multicolumn{1}{l}{} & \multicolumn{1}{l}{} & \multicolumn{1}{l}{} & \multicolumn{1}{l}{} & \multicolumn{1}{l}{} & \multicolumn{1}{l}{} &  \\ 
\hline
\textbf{mFLOT} & 0.41 & 0.52 & 0.65 & 0.55 & 0.49 & 0.43 & 0.65 & 0.60 & 0.72 & \multirow{13}{*}{\textbf{\rotatebox{-90}{RAYSTATION}}} \\
\textbf{sdFLOT} & 0.17 & 0.22 & 0.28 & 0.28 & 0.37 & 0.18 & 0.37 & 0.37 & 0.36 &  \\
\textbf{mdFLOT} & 0.41 & 0.52 & 0.63 & 0.52 & 0.22 & 0.39 & 1.00 & 0.57 & 1.00 &  \\
\textbf{moFLOT} & 1.00 & 1.00 & 1.00 & 1.00 & 1.00 & 1.00 & 1.00 & 1.00 & 1.00 &  \\
\textbf{minFLOT} & 0.08 & 0.14 & 0.15 & 0.18 & 0.17 & 0.22 & 0.20 & 0.17 & 0.17 &  \\
\textbf{maxFLOT} & 1.00 & 1.00 & 1.00 & 1.00 & 1.00 & 1.00 & 1.00 & 1.00 & 1.00 &  \\
\textbf{CFNS$_{5}$} & 0.00 & 0.00 & 0.00 & 0.00 & 0.00 & 0.00 & 0.00 & 0.00 & 0.00 &  \\
\textbf{CFNS$_{10}$} & 0.06 & 0.00 & 0.00 & 0.00 & 0.00 & 0.00 & 0.00 & 0.00 & 0.00 &  \\
\textbf{CFNS$_{25}$} & 0.15 & 0.14 & 0.14 & 0.22 & 0.52 & 0.11 & 0.33 & 0.35 & 0.24 &  \\
\textbf{CFNS$_{50}$} & 0.80 & 0.43 & 0.29 & 0.47 & 0.61 & 0.73 & 0.44 & 0.46 & 0.34 &  \\
\textbf{CFNS$_{75}$} & 0.95 & 0.88 & 0.66 & 0.78 & 0.67 & 0.95 & 0.48 & 0.56 & 0.39 &  \\
\textbf{CFNS$_{90}$} & 0.99 & 0.91 & 0.69 & 0.80 & 0.68 & 0.96 & 0.48 & 0.57 & 0.40 &  \\
\textbf{CFNS$_{95}$} & 0.99 & 0.91 & 0.69 & 0.80 & 0.69 & 0.96 & 0.49 & 0.57 & 0.41 &  \\
\hline
\end{tabular}
\caption{Relative LOT statistics extracted from the plans in the reference dataset.}
\end{table}

\begin{table}
\centering
\begin{tabular}{l|rrr|rrr|rrr|l} 
\hline
\multicolumn{1}{c}{} & \multicolumn{3}{c}{\textbf{Prostate}} & \multicolumn{3}{c}{\textbf{Head and neck}} & \multicolumn{3}{c}{\textbf{Mesothelioma}} & \multicolumn{1}{c}{} \\
\cline{1-10}\arrayrulecolor{black}\cline{11-11}
\multicolumn{1}{l}{} & \multicolumn{1}{l}{\textbf{RP01}} & \multicolumn{1}{l}{\textbf{RP02}} & \multicolumn{1}{l}{\textbf{RP03}} & \multicolumn{1}{l}{\textbf{RP04}} & \multicolumn{1}{l}{\textbf{RP05}} & \multicolumn{1}{l}{\textbf{RP06}} & \multicolumn{1}{l}{\textbf{RP07}} & \multicolumn{1}{l}{\textbf{RP08}} & \multicolumn{1}{l}{\textbf{RP09}} &  \\ 
\hline
\textbf{L$0$NS} & 0.00 & 0.00 & 0.00 & 0.04 & 0.02 & 0.03 & 0.01 & 0.02 & 0.01 & \multirow{13}{*}{\textbf{{\rotatebox{-90}{PRECISION}}}} \\
\textbf{L$1$NS} & 0.26 & 0.28 & 0.22 & 0.27 & 0.23 & 0.27 & 0.22 & 0.23 & 0.26 &  \\
\textbf{L$2$NS} & 0.74 & 0.71 & 0.78 & 0.69 & 0.76 & 0.70 & 0.77 & 0.76 & 0.73 &  \\
\textbf{nCC} & 1.01 & 1.07 & 1.00 & 1.04 & 1.45 & 1.93 & 1.67 & 1.66 & 2.12 &  \\
\textbf{lengthCC} & 8.00 & 7.56 & 9.85 & 9.03 & 10.08 & 7.57 & 10.39 & 9.97 & 7.84 &  \\
\textbf{TA} & 8.13 & 8.21 & 9.89 & 9.97 & 15.81 & 18.08 & 20.30 & 20.15 & 21.27 &  \\
\textbf{fDISC} & 0.01 & 0.07 & 0.01 & 0.12 & 0.47 & 0.74 & 0.57 & 0.55 & 0.76 &  \\
\textbf{CLS} & 0.87 & 0.87 & 0.85 & 0.85 & 0.77 & 0.77 & 0.73 & 0.74 & 0.74 &  \\
\textbf{CLS$_{in}$} & 0.00 & 0.00 & 0.00 & 0.01 & 0.02 & 0.05 & 0.05 & 0.06 & 0.07 &  \\
\textbf{CLS$_{in,disc}$} & 0.03 & 0.03 & 0.03 & 0.08 & 0.04 & 0.07 & 0.08 & 0.10 & 0.10 &  \\
\textbf{CLS$_{in,area}$} & 0.00 & 0.01 & 0.00 & 0.04 & 0.07 & 0.18 & 0.14 & 0.14 & 0.19 &  \\
\textbf{CLS$_{in,area,disc}$} & 0.16 & 0.18 & 0.15 & 0.35 & 0.15 & 0.23 & 0.23 & 0.24 & 0.25 &  \\
\textbf{centroid} & 2.07 & 2.17 & 2.52 & 6.60 & 4.44 & 5.52 & 9.75 & 11.12 & 12.28 &  \\ 
\hline
\multicolumn{1}{l}{} & \multicolumn{1}{l}{} & \multicolumn{1}{l}{} & \multicolumn{1}{l}{} & \multicolumn{1}{l}{} & \multicolumn{1}{l}{} & \multicolumn{1}{l}{} & \multicolumn{1}{l}{} & \multicolumn{1}{l}{} & \multicolumn{1}{l}{} &  \\ 
\hline
\textbf{L$0$NS} & 0.00 & 0.00 & 0.01 & 0.01 & 0.04 & 0.01 & 0.01 & 0.02 & 0.04 & \multirow{13}{*}{\textbf{\rotatebox{-90}{RAYSTATION}}} \\
\textbf{L$1$NS} & 0.24 & 0.23 & 0.21 & 0.23 & 0.33 & 0.23 & 0.23 & 0.25 & 0.34 &  \\
\textbf{L$2$NS} & 0.76 & 0.77 & 0.78 & 0.76 & 0.63 & 0.76 & 0.76 & 0.74 & 0.62 &  \\
\textbf{nCC} & 1.07 & 1.14 & 1.14 & 1.31 & 2.16 & 2.23 & 1.92 & 2.12 & 2.27 &  \\
\textbf{lengthCC} & 8.52 & 8.85 & 9.43 & 10.10 & 5.73 & 8.52 & 9.38 & 8.22 & 5.72 &  \\
\textbf{TA} & 9.20 & 10.34 & 10.99 & 14.17 & 15.89 & 22.39 & 22.43 & 24.27 & 18.82 &  \\
\textbf{fDISC} & 0.06 & 0.14 & 0.14 & 0.27 & 0.71 & 0.92 & 0.69 & 0.77 & 0.78 &  \\
\textbf{CLS} & 0.86 & 0.84 & 0.83 & 0.79 & 0.81 & 0.70 & 0.72 & 0.73 & 0.80 &  \\
\textbf{CLS$_{in}$} & 0.00 & 0.00 & 0.00 & 0.02 & 0.06 & 0.05 & 0.07 & 0.11 & 0.09 &  \\
\textbf{CLS$_{in,disc}$} & 0.02 & 0.03 & 0.03 & 0.06 & 0.08 & 0.06 & 0.10 & 0.14 & 0.12 &  \\
\textbf{CLS$_{in,area}$} & 0.01 & 0.02 & 0.02 & 0.06 & 0.19 & 0.15 & 0.18 & 0.23 & 0.25 &  \\
\textbf{CLS$_{in,area,disc}$} & 0.13 & 0.16 & 0.13 & 0.22 & 0.26 & 0.17 & 0.26 & 0.30 & 0.31 &  \\
\textbf{centroid} & 2.45 & 2.66 & 2.80 & 5.91 & 4.67 & 6.41 & 10.70 & 12.58 & 12.22 &  \\
\hline
\end{tabular}
\caption{Geometrical metrics extracted from the reference dataset.}
\end{table}

\begin{table}
\centering
\begin{tabular}{l|rrr|rrr|rrr|l} 
\hline
\multicolumn{1}{c}{} & \multicolumn{3}{c}{\textbf{Prostate}} & \multicolumn{3}{c}{\textbf{Head and neck}} & \multicolumn{3}{c}{\textbf{Mesothelioma}} & \multicolumn{1}{c}{} \\
\cline{1-10}\arrayrulecolor{black}\cline{11-11}
 & \multicolumn{1}{l}{\textbf{RP01}} & \multicolumn{1}{l}{\textbf{RP02}} & \multicolumn{1}{l}{\textbf{RP03}} & \multicolumn{1}{l}{\textbf{RP04}} & \multicolumn{1}{l}{\textbf{RP05}} & \multicolumn{1}{l}{\textbf{RP06}} & \multicolumn{1}{l}{\textbf{RP07}} & \multicolumn{1}{l}{\textbf{RP08}} & \multicolumn{1}{l}{\textbf{RP09}} &  \\ 
\hline
\textbf{nOC} & 0.25 & 0.25 & 0.31 & 0.29 & 0.46 & 0.46 & 0.54 & 0.52 & 0.52 & \multirow{14}{*}{\textbf{{\rotatebox{-90}{PRECISION}}}} \\
\textbf{PSTV} & 2.53 & 3.07 & 2.73 & 3.78 & 5.55 & 7.59 & 6.96 & 6.90 & 10.69 &  \\
\textbf{EPSTV$_{1,1}$} & 2.53 & 3.07 & 2.73 & 3.78 & 5.55 & 7.59 & 6.96 & 6.90 & 10.69 &  \\
\textbf{EPSTV$_{0,1}$} & 0.58 & 0.80 & 0.77 & 1.83 & 2.20 & 3.19 & 3.18 & 3.22 & 5.55 &  \\
\textbf{EPSTV$_{1,0}$} & 1.94 & 2.27 & 1.96 & 1.94 & 3.35 & 4.40 & 3.79 & 3.68 & 5.14 &  \\
\textbf{LOTV} & 0.94 & 0.93 & 0.95 & 0.96 & 0.93 & 0.91 & 0.95 & 0.95 & 0.91 &  \\
\textbf{ELOTV$_1$} & 0.06 & 0.07 & 0.05 & 0.04 & 0.07 & 0.09 & 0.05 & 0.05 & 0.09 &  \\
\textbf{ELOTV$_3$} & 0.15 & 0.16 & 0.15 & 0.08 & 0.17 & 0.22 & 0.13 & 0.13 & 0.17 &  \\
\textbf{ELOTV$_5$} & 0.21 & 0.22 & 0.21 & 0.11 & 0.23 & 0.29 & 0.17 & 0.18 & 0.21 &  \\
\textbf{MI} & 6.86 & 7.20 & 6.26 & 7.31 & 10.32 & 11.45 & 11.40 & 9.96 & 14.93 &  \\
\textbf{mSI} & 0.43 & 0.36 & 0.40 & 0.11 & 0.26 & 0.29 & 0.17 & 0.19 & 0.16 &  \\
\textbf{mdSI} & 0.53 & 0.44 & 0.51 & 0.12 & 0.30 & 0.34 & 0.19 & 0.23 & 0.16 &  \\
\textbf{sdSI} & 0.18 & 0.18 & 0.20 & 0.08 & 0.17 & 0.15 & 0.08 & 0.09 & 0.06 &  \\
\textbf{MSA} & 0.99 & 0.88 & 0.96 & -0.31 & 0.97 & 0.96 & 1.50 & 0.13 & -0.15 &  \\ 
\hline
\multicolumn{1}{l}{} & \multicolumn{1}{l}{} & \multicolumn{1}{l}{} & \multicolumn{1}{l}{} & \multicolumn{1}{l}{} & \multicolumn{1}{l}{} & \multicolumn{1}{l}{} & \multicolumn{1}{l}{} & \multicolumn{1}{l}{} & \multicolumn{1}{l}{} &  \\ 
\hline
\textbf{nOC} & 0.28 & 0.29 & 0.25 & 0.35 & 0.30 & 0.58 & 0.32 & 0.36 & 0.22 & \multirow{14}{*}{\textbf{\rotatebox{-90}{RAYSTATION}}}\\
\textbf{PSTV} & 2.03 & 2.89 & 3.78 & 4.81 & 6.71 & 5.20 & 7.12 & 7.74 & 8.94 &  \\
\textbf{EPSTV$_{1,1}$} & 2.03 & 2.89 & 3.78 & 4.81 & 6.71 & 5.20 & 7.12 & 7.74 & 8.94 &  \\
\textbf{EPSTV$_{0,1}$} & 0.48 & 0.81 & 1.15 & 2.02 & 2.84 & 1.98 & 3.25 & 3.63 & 4.35 &  \\
\textbf{EPSTV$_{1,0}$}& 1.56 & 2.08 & 2.63 & 2.79 & 3.87 & 3.22 & 3.87 & 4.11 & 4.59 &  \\
\textbf{LOTV} & 0.96 & 0.94 & 0.93 & 0.94 & 0.92 & 0.94 & 0.94 & 0.94 & 0.93 &  \\
\textbf{ELOTV$_1$}  & 0.04 & 0.06 & 0.07 & 0.06 & 0.08 & 0.06 & 0.06 & 0.06 & 0.07 &  \\
\textbf{ELOTV$_3$}& 0.11 & 0.14 & 0.16 & 0.13 & 0.18 & 0.12 & 0.14 & 0.13 & 0.15 &  \\
\textbf{ELOTV$_5$} & 0.15 & 0.19 & 0.23 & 0.18 & 0.23 & 0.16 & 0.20 & 0.18 & 0.19 &  \\
\textbf{MI} & 6.70 & 7.40 & 7.60 & 8.92 & 11.00 & 12.57 & 10.40 & 11.43 & 11.93 &  \\
\textbf{mSI} & 0.27 & 0.38 & 0.39 & 0.20 & 0.18 & 0.21 & 0.20 & 0.16 & 0.16 &  \\
\textbf{mdSI} & 0.28 & 0.47 & 0.50 & 0.24 & 0.18 & 0.25 & 0.23 & 0.19 & 0.16 &  \\
\textbf{sdSI} & 0.15 & 0.20 & 0.27 & 0.11 & 0.13 & 0.14 & 0.09 & 0.07 & 0.06 &  \\
\textbf{MSA} & 0.99 & 0.94 & 1.04 & 0.89 & 1.01 & 0.94 & 1.33 & 0.18 & 0.04 &  \\
\hline
\end{tabular}
\caption{Modulation metrics extracted from the reference dataset.}
\end{table}

\bibliographystyle{ieeetr}


\end{document}